\documentclass[reprint,showpacs,preprintnumbers,amsmath,amssymb,aps,prl,showkeys,nofootinbib,superscriptaddress]{revtex4-1}

\usepackage[utf8]{inputenc}
\usepackage{graphicx}
\usepackage{dcolumn}
\usepackage{bm}

\begin{document}

\title{Noise Spectroscopy with a Quantum Gas}

\author{P. Federsel}
\affiliation{Physikalisches Institut, Eberhardt-Karls-Universität Tübingen, D-72076 Tübingen, Germany}
\author{C. Rogulj}
\affiliation{Physikalisches Institut, Eberhardt-Karls-Universität Tübingen, D-72076 Tübingen, Germany}
\author{T. Menold}
\affiliation{Physikalisches Institut, Eberhardt-Karls-Universität Tübingen, D-72076 Tübingen, Germany}
\author{Z. Darázs}
\affiliation{Institute for Solid State Physics and Optics, Wigner Research Centre for Physics,\\Hungarian Academy of Sciences, P.O. Box 49, H-1525
Budapest, Hungary}
\author{P. Domokos}
\affiliation{Institute for Solid State Physics and Optics, Wigner Research Centre for Physics,\\Hungarian Academy of Sciences, P.O. Box 49, H-1525
Budapest, Hungary}
\author{A. Günther}
\email[corresponding author: ]{a.guenther@uni-tuebingen.de}
\affiliation{Physikalisches Institut, Eberhardt-Karls-Universität Tübingen, D-72076 Tübingen, Germany}
\author{J. Fortágh}
\affiliation{Physikalisches Institut, Eberhardt-Karls-Universität Tübingen, D-72076 Tübingen, Germany}

\date{\today}

\begin{abstract}
We report on the spectral analysis and the local measurement of intensity correlations of microwave fields using ultra cold quantum gases. The fluctuations of the electromagnetic field induce spin flips in a magnetically trapped quantum gas and generate a multi-mode atomlaser. The output of the atomlaser is measured with high temporal resolution on the single atom level, from which the spectrum and intensity correlations of the generating microwave field are reconstructed. We give a theoretical description of the atomlaser output and its correlations in response to resonant microwave fields and verify the model with measurements on an atom chip. The measurement technique is applicable for the local analysis of classical and quantum noise of electromagnetic fields, for example on chips, in the vicinity of quantum electronic circuits.
\end{abstract}

\maketitle

Fluctuations and noise play an important role in our fundamental understanding of classical and quantum systems. In the famous Hanbury-Brown and Twiss experiment intensity fluctuations have been used to determine coherence properties of chaotic light \cite{brown1956}. Similar effects have been observed for massive particles, such as bosons and fermions, showing bunching \cite{Schellekens2005} and anti-bunching \cite{kiesel2002} in the particle correlations. Transport phenomena in solid-state quantum devices, such as single electron transport through quantum dots \cite{andergassen2010} or ballistic transport in graphene \cite{novoselov2005}, are well characterized by the electron counting statistics and the corresponding field noise. This becomes especially important, as novel materials such as artificial honeycomb crystals \cite{beugeling2015} predict quantum effects in the electron transport even at room temperature, due to the formation of topological protected states \cite{kalesaki2014}. Such quantum transport phenomena might be measured by means of a recently proposed quantum galvanometer \cite{Kalman2012}, in which the low frequency current noise of a nano-device is coherently coupled to an atomic quantum gas and analyzed via state selective single atom detection.
 
Here, we demonstrate the basic operation of the quantum galvanometer and extend it to quantum correlation measurements. Using a Bose-Einstein condensate, we coherently probe artificial, low frequency magnetic field fluctuations (noise) by shifting them electronically into the microwave (mw) regime, close to an atomic resonance. These fluctuations, generate a multi-mode atomlaser, with an output directly connected to the original field fluctuations. Using a sensitive detector, we analyze this output on the single atom level and show, how the power spectral density and the intensity correlations of the microwave field can be reconstructed. We give a theoretical description for the output of the multi-mode atomlaser, including decoherence effects.

\begin{figure}[t]
    \includegraphics{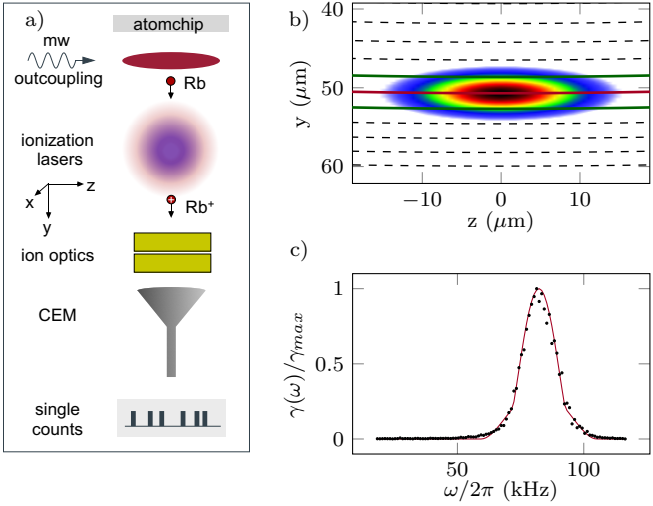}
    \caption{(a) Cold atom spectrometer (not to scale) consisting of a magnetically trapped Bose-Einstein condensate and an ionization based single atom detector. (b) The mw couples atoms at resonance surfaces, given by equipotential surfaces of the atomic Zeeman potential, i.e. magnetic iso-field lines (dashed lines). Due to gravity, the BEC is displaced from the magnetic trap center and the resonance surfaces become nearly plane. Without amplitude modulation, the mw-carrier couples atoms from a single resonance surface (red line) with a position given via $\omega_c$. Amplitude modulation at a single frequency generates sidebands to the carrier, and outcoupling from two resonance surfaces (green lines) (c) Normalized spectral response $\gamma(\omega)/\gamma_{max}$ of a BEC to a single mw-frequency (black dots) and model function (red line).}
    \label{fig:schematics}
\end{figure}

\paragraph{Experimental setup:}
The experiment is illustrated in Fig. \ref{fig:schematics}a. Using an atom-chip based cold atom apparatus \cite{Guenther2005a}, we prepare Bose-Einstein condensates and thermal ensembles of $^{87}$Rb atoms in the $\left|5S_{1/2},F=2,m_F=2\right>$ ground state. The atoms are magnetically trapped in a harmonic configuration with trap frequencies $\omega_{\left(x,y,z\right)}=2\pi\times\left(85, 70, 16\right)$Hz and offset field $B_{z,\mbox{\tiny off}}\approx0.93$G. If this cloud is exposed to resonant microwave radiation, spin flips to the anti-trapped $\left|5S_{1/2},F=1,m_F=1\right>$ state occur. Here, we irradiate microwaves of various spectra to demonstrate the measurement of noise spectra and correlations. In particular, we apply amplitude modulation to a microwave carrier at $\omega_c\approx2\pi\times 6.834$GHz with a variable function $A\left(t\right)$ in the $kHz$ regime. Here, $A(t)$ mimics the low frequency field noise, which in the quantum galvanometer case is intrinsically (via a mechanical oscillation of the current driven nano device) mixed up to atomic transition frequencies in the MHz regime \cite{Kalman2012}. The magnetic coupling field at the position of the atoms is then given by $|\vec{B}\left(t\right)|=A\left(t\right)\cdot B_0\cdot\cos\left(\omega_ct+\phi\right)$ with $B_0$ and $\varphi$ being amplitude and phase of the microwave carrier. The amplitude modulation produces sidebands to the carrier frequency $\omega_c$. Each frequency component of the microwave addresses atoms at different resonance surfaces of the trap (see Fig. \ref{fig:schematics}b). Adjusting $\omega_c$ and $A(t)$, individual or multiple regions of the BEC can be addressed at the same time. Spin flipped atoms leave the trap and are detected with single atom resolution and $\sim19\%$ efficiency, using a multi-photon ionization process and subsequent ion counting \cite{Stibor2010}.

\paragraph{Atomic response:}
A monochromatic microwave of amplitude $B_0$ and frequency $\omega$ couples out a coherent atomlaser beam from a BEC. In the limit of weak outcoupling, losses are negligible and the outcoupling rate is $\Gamma\left(\omega\right)  = \gamma\left(\omega\right)\cdot B_0^2$. The spectral response $\gamma(\omega)$ can be measured as shown in Fig. \ref{fig:schematics}c), yielding a frequency width largely increased due to gravity \cite{Federsel2015a,Kalman2016}.

Using a wavelike approach, where the phase of the atomlaser is locked to the microwave field \cite{Vainio2006,Esslinger2000} and different atomlasers may interfere with each other \cite{Bloch2000,Bourdel2006}, the outcoupling rate can be extended to arbitrary fields $B\left(t\right)$
\begin{flalign}
\Gamma\left(t\right)&\!=\!\left|\frac{1}{\pi}\int\limits_{0}^{\infty}\!\tilde{B}\left(\omega \right)\sqrt{\gamma\left(\omega \right)}e^{i\omega t}d\omega\right|^2\ast V\left(t\right)\label{rate_matterwave}\\
&=\frac{1}{\pi^2}\int\limits_{-\infty}^{\infty}e^{i\Delta\omega t}R_{\xi\xi}\left(\Delta\omega\right)\tilde{V}\left(\Delta\omega\right)d\Delta\omega\label{rate_spectral_osc}
\end{flalign}
with $\ast$ being the convolution, $\tilde{B}\left(\omega\right)$ the fourier transform\footnote{The fourier transform $\mathcal{F}$ is defined by $\mathcal{F}\left(B\left(t\right)\right)=\int\limits_{-\infty}^{\infty}B\left(t\right)e^{-i\omega t}dt=\tilde{B}\left(\omega\right)$} of $B(t)$ and $R_{\xi\xi}$ the autocorrelation of  $\xi\left(\omega\right)=\tilde{\Theta}\left(\omega\right)\tilde{B}\left(\omega\right)\sqrt{\gamma\left(\omega\right)}$ with the Heaviside function $\tilde{\Theta}\left(\omega\right)$. The visibility function $\tilde{V}\left(\Delta\omega\right)=\mathcal{F}\left(V\left(t\right)\right)$ has been included to account for the detector's finite temporal resolution and decoherence effects, which may arise from the atoms' finite coherence length. $\tilde{V}$ is expected to be symmetric with $\tilde{V}(|\Delta\omega|)\leq 1$ and $\tilde{V}(0)=1$. The time-averaged countrate $\left<\Gamma(t)\right>$ can be found from Eq. \ref{rate_matterwave} in the limit $V(t)\rightarrow const$. This corresponds to the incoherent case with $V(\Delta\omega\neq 0)=0$, for which the outcoupling rate becomes time independent and Eq. \ref{rate_spectral_osc} yields
\begin{equation}
\left<\Gamma(t)\right>=\frac{1}{\pi^2}R_{\xi\xi}(0)=\frac{1}{\pi^2}\int\limits_{0}^{\infty} S_{BB}\left(\omega\right) \gamma\left(\omega\right) d\omega
\label{eq:gamma_tot}
\end{equation}
with the power spectral density $S_{BB}\left(\omega\right)=|\tilde{B}(\omega)|^2$.

\paragraph{Noise analysis:}
Spectroscopic information about the local magnetic field at the atomic position can be gained from analyzing either the time dependent outcoupling $\Gamma(t)$ or its time-average $\left<\Gamma\right>$. This defines two possible measurement modes:

The \textit{spectrometer mode} concentrates on measuring time-averaged countrates, which are according to Eq. \ref{eq:gamma_tot} independent from the visibility function $\tilde V$. In this mode, information about the power spectral density can be gained, by making the BEC sensitive to different parts of the spectrum, thus measuring $\left<\Gamma\right>$ while tuning the difference $\delta\omega$ between the center of the spectral response and the microwave spectrum. This can be reached by either shifting $\gamma$ via the magnetic offset field or by shifting $S_{BB}$ via the carrier frequency. The mean outcoupling rate from Eq. \eqref{eq:gamma_tot} then reads
\begin{align}
\left<\Gamma\left(t\right)\right>\left(\delta\omega\right) & =\frac{1}{\pi^2}\int\limits_{0}^{\infty} S_{BB}\left(\omega+\delta\omega\right) \gamma\left(\omega\right) d\omega\\
&=\frac{1}{2\pi^2}\left(S_{BB}(\omega)\ast\gamma(\omega)\right)\left(\delta\omega\right)\label{eq:spectrum_convolution}
\end{align}
Here we used $\gamma(-\omega)=\gamma(\omega)$ and $S_{BB}(-\omega)=S_{BB}(\omega)$ which is valid for classical fields. If the spectral response function is known, the power spectral density can be reconstructed via a deconvolution. It is then a direct measure for the power spectrum $S_{AA}(\omega)=\left|\tilde{A}(\omega)\right|^2$ of the low frequency noise, since $S_{BB}(\omega)=\pi^2 B_0^2 S_{AA}(\omega-\omega_c)$.

The \textit{correlator mode} concentrates on analyzing the time dependent signals $\Gamma(t)$ and the corresponding second-order correlation function $g^{(2)}(\tau)=\left<\Gamma(t)\Gamma(t+\tau)\right>/\left(\left<\Gamma(t)\right>\left<\Gamma(t+\tau)\right>\right)$ reads
\begin{eqnarray}
g^{(2)}(\tau)&=&\frac{4\pi^2}{R_{\xi\xi}(0)}\mathcal{F}^{-1}\left(\left|R_{\xi\xi}(\Delta\omega)\tilde V(\Delta\omega)\right|^2\right)\label{eq:g22}\\
&=&ACF\left(\left|\xi(t)\right|^2\right)\ast V(t)\ast V(t)\label{eq:g23}
\end{eqnarray}
with $\xi(t)=\mathcal{F}^{-1}(\tilde\xi(\omega))(t)$ and ACF being the autocorrelation function. Using the theory of analytic signals \cite{Yarlagadda2010}, one can show that the envelope of any real valued function $f(t)$ can be calculated via $env\left(f(t)\right)=\left|\Theta(t)\ast f(t)\right|$ \cite{Yarlagadda2010} with $\Theta(t)$ being the inverse Fourier transform of the Heaviside function. Following this, one finds
\begin{eqnarray}
\left|\xi(t)\right|^2 &=& \left|\Theta\left(t\right)\ast \mathcal{F}^{-1}\left(\tilde{B}\left(\omega\right)\sqrt{\gamma\left(\omega\right)}\right)\right|^2\\
&=& env\left(B(t)\ast\mathcal{F}^{-1}\left(\sqrt{\gamma(\omega)}\right)\right)^2 = I_{fil}(t)
\end{eqnarray}
with $I_{fil}(t)$ being the spectrally filtered microwave intensity and $env\left(B(t)\right)\propto \left|A(t)\right|$. Measuring $g^{(2)}$ will thus directly unveil intensity correlations of the radiation field within the bandwidth of the quantum gas.

The correlator mode requires knowledge of either $\tilde V(\Delta\omega)$ or $V(t)$. The former is most easily measured by outcoupling with two frequencies, $B=B_0(\cos(\omega_1 t+\varphi_1) + \cos(\omega_2 t + \varphi_2))$. Choosing the frequencies such that $\gamma\left(\omega_1\right)=\gamma\left(\omega_2\right)$ Eq. \ref{rate_spectral_osc} becomes
\begin{equation}
\Gamma\left(t\right)\sim \left(1+\tilde{V}\left(\left|\Delta\omega\right|\right)\cos\left(\Delta\omega t+\Delta\phi\right)\right)\label{eq:rate_2freq}
\end{equation}
and
\begin{equation}
g^{(2)}(\tau)=1+\frac{\tilde V(\left|\Delta\omega\right|)^2}{2}\cos\left(\Delta\omega \tau\right)\label{eq:g2}
\end{equation}
with $\Delta\omega=\omega_2-\omega_1$ and $\Delta\phi=\phi_2-\phi_1$. The visibility function is thus directly connected to the interference contrast of the two beating atomlasers and can be measured by varying $\Delta\omega$. $V(t)$ on the other hand can be measured in the limit of short mw-pulses and spectral response functions much broader than the bandwidth of $\tilde{V}\left(\Delta\omega\right)$. In this case $\tilde B(\omega),\gamma(\omega)\rightarrow const$ and Eq. \ref{rate_matterwave} becomes
\begin{equation}
\Gamma_\sqcap(t)\sim\delta(t)\ast V(t) = V(t)
\end{equation}
For spectral response functions of finite width and mw-pulses of finite length, however, Eq. \ref{rate_matterwave} reads
\begin{equation}
\Gamma_\sqcap(t)\sim\left|\mathcal{F}^{-1}\left(\tilde\theta(\omega)\tilde B(\omega)\sqrt{\gamma(\omega)}\right)(t)\right|^2\ast V(t)\label{eq:pulse}
\end{equation}
causing a slight broadening of the measured pulse response $\Gamma_\sqcap(t)$ with respect to $V(t)$.

Using our single atom detector, both $\left<\Gamma(\delta\omega)\right>$ and $\Gamma(t)$ can be measured in-situ and in real-time. In practise $\left<\Gamma(\delta\omega)\right>$ (spectrometer mode) is best used for measuring broad-band power spectral densities with a resolution limited by the atoms' spectral response. Spectral information within the atomic bandwidth, however, can be obtained in form of intensity correlations, from measuring $\Gamma(t)$ or $g^{(2)}(\tau)$ (correlator mode) for a fully coherent object like a BEC. In this mode, the bandwidth of the spectral response sets an upper limit for the fastest detectable correlations.

\begin{figure}[t]
	\includegraphics{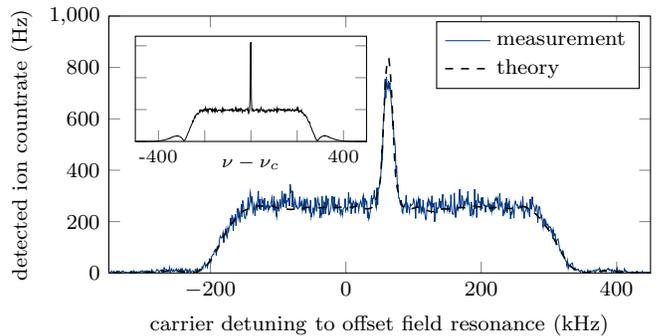}
    \caption{Spectrometer mode. Mean ion countrate for different microwave carrier detunings $\delta\omega$. Inset: Power spectrum of the original microwave as measured with a spectrum analyzer. Experimental data (blue lines) are shown together with the theory form Eq. \ref{eq:spectrum_convolution} (dashed line). The form of the model function from \ref{fig:schematics}c) was used to calculate the convolution. The amplitude was calculated from the theory of \cite{Federsel2015a} with $N=8900$ and $T=33.75$nK.}
    \label{fig:expmeasmode1}
\end{figure}

\paragraph{Measurements:}
To demonstrate the spectrometer mode, we generate a broad-band mw-spectrum via amplitude modulation of a microwave carrier. As source for the amplitude modulation $A\left(t\right)$ we use a low-pass filtered noise diode with cutoff at around $200$kHz. The resulting power spectral density is shown in the inset of Fig. \ref{fig:expmeasmode1}.
As atomic probe we use a BEC with 8900 atoms and expose it to the mw-spectrum which is shifted within 700ms by sweeping the carrier frequency with $1.4$kHz/ms. Outcoupled atoms are photoionized by two overlapping laser beams \cite{Stibor2010} of about $50\mu$m waist, which are positioned $370 \mu m$ below the atomic cloud (see Fig. \ref{fig:schematics}a). The number of detected ions per sweep is about 50 and their arrival times can be mapped to the carrier detuning. Repeating the measurement 1500 times, the ion countrate can be deduced, as shown in Fig. \ref{fig:expmeasmode1}. The measured data is in agreement with the theory from Eq. \ref{eq:spectrum_convolution}, which has been derived by convolving the microwave spectrum with the spectral response function $\gamma\left(\omega\right)$ from Fig. \ref{fig:schematics}c.

\begin{figure}
	\includegraphics{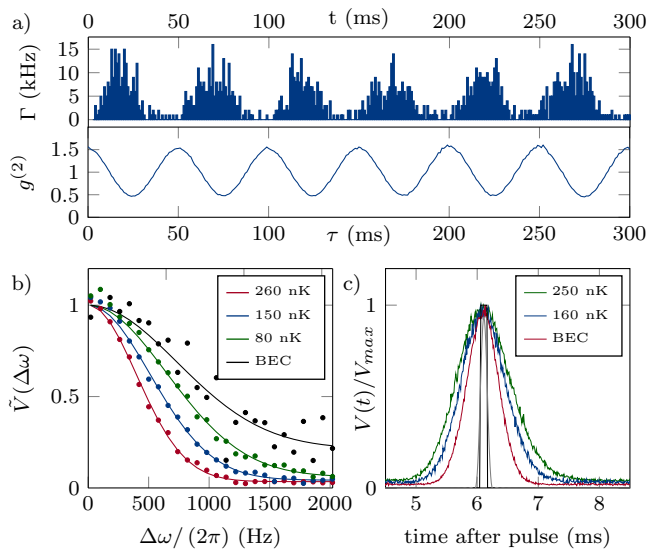}
    \caption{a) Ion countrate (upper panel) and $g^{(2)}$-correlation function (lower panel) of two interfering atomlasers with $\Delta\omega=2\pi\times 20$Hz. A fit of the theory from Eq. \ref{eq:rate_2freq} and \ref{eq:g2} allows for extracting the visibility $\tilde V(\Delta\omega)$. b) Visibility functions of thermal clouds and BECs, as derived from two interfering atomlaser with variable $\Delta\omega$. Gaussian model functions have been fitted, to deduce the coherence frequency $\sigma$. The fits yield $\sigma=2\pi\times 420$, $540$, $700$, $960$Hz for $260$nK, $150$nK, partially condensed and fully condensed clouds, respectively. c) Temporal ion distribution for pulsed outcoupling from thermal clouds and a BEC with $40\times 10^3$ atoms. Ion distributions are peaked at around $6$ms, which is the mean delay between outcoupling and detection. The solid black line shows the initial pulse of $120\mu$s width. The gray line shows $\left|\mathcal{F}^{-1}\left(\tilde\theta(\omega)\tilde B(\omega)\sqrt{\gamma(\omega)}\right)(t)\right|^2$, which is according to Eq. \ref{eq:pulse} the expected countrate in the limit of full coherence, calculated for a spectral response of 20kHz width.}
    \label{fig:2frequency}
\end{figure}

To interpret time-dependent outcoupling rates, the visibility function $\tilde{V}\left(\Delta\omega\right)$ has to be determined. This can be done by using $A\left(t\right)=A_0\cdot\sin\left(\Delta\omega t/2+\phi\right)$ as amplitude modulation, generating two frequencies at $\omega_{1/2}=\omega_c\pm\Delta\omega/2$. Setting $\omega_c$ to the center of the spectral response assures $\gamma(\omega_1)\approx\gamma(\omega_2)$ and thus according to Eq. \ref{eq:rate_2freq} and \ref{eq:g2} a direct measure of $V(|\Delta\omega|)$. Figure \ref{fig:2frequency}a shows the resulting countrate and the corresponding correlation function for a thermal cloud of 260nK and $\Delta\omega=2\pi\times 20$Hz. Both signals show a clear oscillation at frequency $\Delta\omega$, with the signal-to-noise ratio being much higher in the $g^{(2)}$-analysis \cite{Note1}. Here, the two atomlasers are fully coherent, because the separation between the two outcoupling surfaces $\Delta z=\lambda\hbar/(mg)\Delta\omega$ amounts only $6.2$nm \cite{Federsel2015a}, which is well below the atoms' thermal de Broglie wavelength. Increasing $\Delta\omega$ and thus $\Delta z$, will lead to a reduced visibility. Fig. \ref{fig:2frequency}b shows the measured visibilities for different $\Delta\omega$ and clouds of different temperatures. Within each data set, the visibility drops on a characteristic frequency scale $\sigma$, which we deduce by fitting a gaussian model function $V(\Delta\omega)=\exp\left(-\Delta\omega^2/2\sigma^2\right)$ to the data. As expected, thermal clouds of reduced temperature show increased values of $\sigma$, which correspond to larger de Broglie wavelengths. However, the visibility is not influenced from the atoms' spatial coherence only, but also from the detector's finite temporal resolution. This is best seen in the BEC data, where we would expect full coherence within the spectral response.
The measurement, however, shows a reduced visibility of $\sigma\approx 1$kHz, which we attribute to the temporally delocalized ionization process. Due to the rather large beam profile of the ionization lasers, ionization occurs in a time window of $\sim0.6$ms.
Instead $\tilde{V}(\Delta\omega)$, we can measure $V(t)$ via pulsed outcoupling with $A(t)=A_0 \sum_n \sqcap(t-nT,\Delta t)$ and $\sqcap(t,\Delta t)=\tilde{\Theta}(t+\Delta t/2)-\tilde{\Theta}(t-\Delta t/2)$. This corresponds to a microwave pulse sequence, with periodicity $T=12$ms and pulse width $\Delta t=120\mu$s. Figure \ref{fig:2frequency}c shows the resulting ion distributions for different cloud temperatures, alongside the (shifted) initial pulse and the spectral response broadened pulse, resulting from Eq. \ref{eq:pulse} in the limit of full coherence $\tilde V=1$. Both pulses are sufficiently short, such that the measured ion distributions give the approximate form of $V\left(t\right)$. As expected, $V(t)$ shows a clear temperature dependence, with decreasing temporal width for decreasing cloud temperatures. However, the spatially full coherent BEC, does not approach the spectral response broadened pulse, but shows a pulse width of $\sim0.6$ms, which is mainly given by the time uncertainty of ionization.

\begin{figure}[htbp]
  \vspace{.5cm}
  \includegraphics{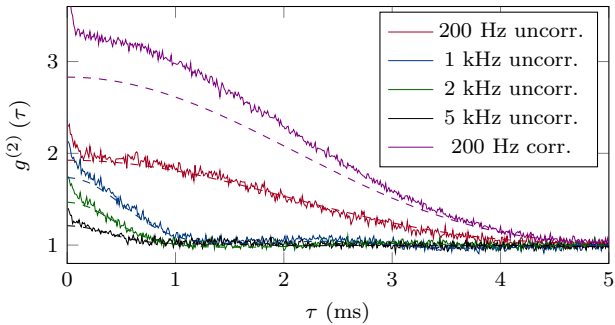}
  \caption{Correlator mode. Second order correlations of the ion arrival times, as measured for a BEC with $40\times 10^3$ atoms while irradiating a phase correlated (corr.) or uncorrelated (uncorr.) microwave noise of variable bandwidth ($200$Hz, $1$kHz, $2$kHz and $5$kHz). The dashed lines show the theory from Eq. \ref{eq:g23}. The increase of the $g^{(2)}$ correlation function on small timescales $\tau<100\mu$s is due to laser intensity noise on the ionization laser, which was verified from a photodiode based intensity measurement and subsequent correlation analysis.}
\label{fig:differentbws}
\end{figure}

Having the visibility function at hand, we use Bose-Einstein condensates to demonstrate the correlator mode. Therefore, we investigate narrow-band microwave noise spectra with adjustable bandwidths, generated via a modulation $A\left(t\right)=A_0\sum_{\nu_i\in\left[\nu_1,\nu_2\right]} \sin(2\pi\nu_it+\phi_i)$, in which the frequencies are chosen in 1Hz steps and the angles $\phi_i$ are chosen randomly. The outcoupling was adjusted such that the atomic cloud is sensitive to the right sideband only, by choosing $\nu_1$ much bigger than the bandwidth of the condensate. Using this artificial spectrum, we measure the time-dependent outcoupling rate from which we deduce the $g^{(2)}$ correlations. Figure \ref{fig:differentbws} shows $g^{2}(\tau)$, as measured with a BEC of $40\times 10^3$ atoms, for bandwidths $\Delta\nu=\nu_2-\nu_1$ ranging from $200$Hz up to $5$kHz. On short timescales all measurements show clear correlations with $g^{(2)}>1$. These correlations decay on timescales on the order of the inverse mw-bandwidth, until the system becomes fully uncorrelated. According to Eq. \ref{eq:g23}, $g^{(2)}$ is directly connected to the mw intensity correlations, which are due to the finite bandwidth of the mw noise. Within this bandwidth, all atomlasers interfere mutually, resulting in multiple overlapping two-beam interferences. As the phase information is lost in the correlation analysis (see Eq. \ref{eq:g2}), overlapping $g^{(2)}$-functions of different frequencies will peak at $\tau=0$ and become uncorrelated on a timescale proportional to the inverse bandwidth. The theory shows good agreement with the experiment, once $V(t)$ is included. In the case of randomly chosen phases $\phi_i$, the maximum correlation value amounts $g^{(2)}(0)=2$, similar to chaotic light in the Hanbury-Brown and Twiss experiment. The multi-frequency outcoupling with random phases thus generates a pseudothermal atom distribution \cite{Oettl2005}, which is expected to show bunching for bosons at $\tau=0$. However, $g^{(2)}(0)$ drops if the bandwidth of the microwave noise extends that of the visibility function. For the BEC measurements in Fig. \ref{fig:differentbws}, the latter is mainly limited by the $0.6$ms timing resolution of the ionization process, leading to decreasing values $g^{(2)}(0)$ for noise bandwidths larger than $1.7$kHz. Using phase-correlated noise arbitrary correlation values can be generated. The purple dataset in Fig. \ref{fig:differentbws} shows the particle correlations for phase-correlated noise with $\nu_1=1$Hz, $\nu_2=100$Hz and the carrier frequency set to the center of the cloud. This way, the condensate becomes sensitive to both sidebands, which have a fixed phase relation to each other. The measurement shows a maximum correlation value of $g^{(2)}\left(0\right)>2$, as expected from theory. The deviation of experiment from theory in the phase-correlated case is likely due to additionally induced dynamics, which may result from the local depletion of the condensate for sufficient high outcoupling rates. With the fluctuations in the outcoupling rate becoming stronger and stronger for increasing values of $g^{(2)}(0)$, such processes are more likely to occur.

\paragraph{Outlook:}
In conclusion, we demonstrated the local measurement of the spectrum and correlations of a microwave field using ultracold atomic quantum gases. The fluctuations of the field have been transferred onto an atomlaser, whose output is measured with single atom resolution. Analyzing the statistics of the atomlaser we reconstructed the power spectral density and the intensity correlations of the fluctuating electromagnetic field. While power spectral densities can be measured with resolutions down to the bandwidth of the atoms' spectral response, intensity correlations are only limited by the linewidth of the atomic transition. In its current realization our detection scheme features a local sensitivity of few $10pT/\sqrt{Hz}$ [11] corresponding to a flux sensitivity in the $\mu\Phi_0/\sqrt{Hz}$ regime, comparable to commercial squid magnetometers. In the future our scheme might be applicable to study non-classical noise \cite{Jing2000,Jing2001,Haine2005} either via Raman induced outcoupling with squeezed light or by measuring field noise emitted from quantum electronic circuits, such as quantum dots or single walled carbon nanotubes. Quantum effects will then show up either via anti-bunching in the correlation mode or by asymmetries in the emission and absorption low frequency noise spectrum, which can be measured as sidebands on the carrier frequency close to an atomic resonance \cite{Nazarov2009}. A quantum galvanometer \cite{Kalman2012} comes thus in direct reach, allowing to investigate quantum transport phenomena. Due to the $\mu$m-size of the quantum gas such field noise could be measured in the near-field, i.e. on length-scales much smaller than the wavelength of the radiation field.

We acknowledge financial support from the DFG SFB TRR21, FET-Open Xtrack Project HAIRS, the Hungarian Academy of Sciences (LP2011-016), and the NKFIH (K 115624). We thank O. Kalman and T. Kiss for helpful discussions.


\begin{thebibliography}{23}
\makeatletter
\providecommand \@ifxundefined [1]{
 \@ifx{#1\undefined}
}
\providecommand \@ifnum [1]{
 \ifnum #1\expandafter \@firstoftwo
 \else \expandafter \@secondoftwo
 \fi
}
\providecommand \@ifx [1]{
 \ifx #1\expandafter \@firstoftwo
 \else \expandafter \@secondoftwo
 \fi
}
\providecommand \natexlab [1]{#1}
\providecommand \enquote  [1]{``#1''}
\providecommand \bibnamefont  [1]{#1}
\providecommand \bibfnamefont [1]{#1}
\providecommand \citenamefont [1]{#1}
\providecommand \href@noop [0]{\@secondoftwo}
\providecommand \href [0]{\begingroup \@sanitize@url \@href}
\providecommand \@href[1]{\@@startlink{#1}\@@href}
\providecommand \@@href[1]{\endgroup#1\@@endlink}
\providecommand \@sanitize@url [0]{\catcode `\\12\catcode `\$12\catcode
  `\&12\catcode `\#12\catcode `\^12\catcode `\_12\catcode `\%12\relax}
\providecommand \@@startlink[1]{}
\providecommand \@@endlink[0]{}
\providecommand \url  [0]{\begingroup\@sanitize@url \@url }
\providecommand \@url [1]{\endgroup\@href {#1}{\urlprefix }}
\providecommand \urlprefix  [0]{URL }
\providecommand \Eprint [0]{\href }
\providecommand \doibase [0]{http://dx.doi.org/}
\providecommand \selectlanguage [0]{\@gobble}
\providecommand \bibinfo  [0]{\@secondoftwo}
\providecommand \bibfield  [0]{\@secondoftwo}
\providecommand \translation [1]{[#1]}
\providecommand \BibitemOpen [0]{}
\providecommand \bibitemStop [0]{}
\providecommand \bibitemNoStop [0]{.\EOS\space}
\providecommand \EOS [0]{\spacefactor3000\relax}
\providecommand \BibitemShut  [1]{\csname bibitem#1\endcsname}
\let\auto@bib@innerbib\@empty

\bibitem [{\citenamefont {Brown}\ and\ \citenamefont
  {Twiss}(1956)}]{brown1956}
  \BibitemOpen
  \bibfield  {author} {\bibinfo {author} {\bibfnamefont {R.~H.}\ \bibnamefont
  {Brown}}\ and\ \bibinfo {author} {\bibfnamefont {R.}~\bibnamefont {Twiss}},\
  }\href@noop {} {\bibfield  {journal} {\bibinfo  {journal} {Nature (London)}\
  }\textbf {\bibinfo {volume} {178}},\ \bibinfo {pages} {1046} (\bibinfo {year}
  {1956})}\BibitemShut {NoStop}
\bibitem [{\citenamefont {Schellekens}\ \emph {et~al.}(2005)\citenamefont
  {Schellekens}, \citenamefont {Hoppeler}, \citenamefont {Perrin},
  \citenamefont {Gomes}, \citenamefont {Boiron}, \citenamefont {Aspect},\ and\
  \citenamefont {Westbrook}}]{Schellekens2005}
  \BibitemOpen
  \bibfield  {author} {\bibinfo {author} {\bibfnamefont {M.}~\bibnamefont
  {Schellekens}}, \bibinfo {author} {\bibfnamefont {R.}~\bibnamefont
  {Hoppeler}}, \bibinfo {author} {\bibfnamefont {A.}~\bibnamefont {Perrin}},
  \bibinfo {author} {\bibfnamefont {J.~V.}\ \bibnamefont {Gomes}}, \bibinfo
  {author} {\bibfnamefont {D.}~\bibnamefont {Boiron}}, \bibinfo {author}
  {\bibfnamefont {A.}~\bibnamefont {Aspect}}, \ and\ \bibinfo {author}
  {\bibfnamefont {C.~I.}\ \bibnamefont {Westbrook}},\ }\href
  {http://www.sciencemag.org/content/310/5748/648.short} {\bibfield  {journal}
  {\bibinfo  {journal} {Science}\ }\textbf {\bibinfo {volume} {310}},\ \bibinfo
  {pages} {648} (\bibinfo {year} {2005})}\BibitemShut {NoStop}
\bibitem [{\citenamefont {Kiesel}\ \emph {et~al.}(2002)\citenamefont {Kiesel},
  \citenamefont {Renz},\ and\ \citenamefont {Hasselbach}}]{kiesel2002}%
  \BibitemOpen
  \bibfield  {author} {\bibinfo {author} {\bibfnamefont {H.}~\bibnamefont
  {Kiesel}}, \bibinfo {author} {\bibfnamefont {A.}~\bibnamefont {Renz}}, \ and\
  \bibinfo {author} {\bibfnamefont {F.}~\bibnamefont {Hasselbach}},\
  }\href@noop {} {\bibfield  {journal} {\bibinfo  {journal} {Nature (London)}\
  }\textbf {\bibinfo {volume} {418}},\ \bibinfo {pages} {392} (\bibinfo {year}
  {2002})}\BibitemShut {NoStop}
\bibitem [{\citenamefont {Andergassen}\ \emph {et~al.}(2010)\citenamefont
  {Andergassen}, \citenamefont {Meden}, \citenamefont {Schoeller},
  \citenamefont {Splettstoesser},\ and\ \citenamefont
  {Wegewijs}}]{andergassen2010}
  \BibitemOpen
  \bibfield  {author} {\bibinfo {author} {\bibfnamefont {S.}~\bibnamefont
  {Andergassen}}, \bibinfo {author} {\bibfnamefont {V.}~\bibnamefont {Meden}},
  \bibinfo {author} {\bibfnamefont {H.}~\bibnamefont {Schoeller}}, \bibinfo
  {author} {\bibfnamefont {J.}~\bibnamefont {Splettstoesser}}, \ and\ \bibinfo
  {author} {\bibfnamefont {M.}~\bibnamefont {Wegewijs}},\ }\href@noop {}
  {\bibfield  {journal} {\bibinfo  {journal} {Nanotechnology}\ }\textbf
  {\bibinfo {volume} {21}},\ \bibinfo {pages} {272001} (\bibinfo {year}
  {2010})}\BibitemShut {NoStop}
\bibitem [{\citenamefont {Novoselov}\ \emph {et~al.}(2005)\citenamefont
  {Novoselov}, \citenamefont {Geim}, \citenamefont {Morozov}, \citenamefont
  {Jiang}, \citenamefont {Katsnelson}, \citenamefont {Grigorieva},
  \citenamefont {Dubonos},\ and\ \citenamefont {Firsov}}]{novoselov2005}
  \BibitemOpen
  \bibfield  {author} {\bibinfo {author} {\bibfnamefont {K.}~\bibnamefont
  {Novoselov}}, \bibinfo {author} {\bibfnamefont {A.~K.}\ \bibnamefont {Geim}},
  \bibinfo {author} {\bibfnamefont {S.}~\bibnamefont {Morozov}}, \bibinfo
  {author} {\bibfnamefont {D.}~\bibnamefont {Jiang}}, \bibinfo {author}
  {\bibfnamefont {M.}~\bibnamefont {Katsnelson}}, \bibinfo {author}
  {\bibfnamefont {I.}~\bibnamefont {Grigorieva}}, \bibinfo {author}
  {\bibfnamefont {S.}~\bibnamefont {Dubonos}}, \ and\ \bibinfo {author}
  {\bibfnamefont {A.}~\bibnamefont {Firsov}},\ }\href@noop {} {\bibfield
  {journal} {\bibinfo  {journal} {Nature (London)}\ }\textbf {\bibinfo {volume}
  {438}},\ \bibinfo {pages} {197} (\bibinfo {year} {2005})}\BibitemShut
  {NoStop}
\bibitem [{\citenamefont {Beugeling}\ \emph {et~al.}(2015)\citenamefont
  {Beugeling}, \citenamefont {Kalesaki}, \citenamefont {Delerue}, \citenamefont
  {Niquet}, \citenamefont {Vanmaekelbergh},\ and\ \citenamefont
  {Smith}}]{beugeling2015}
  \BibitemOpen
  \bibfield  {author} {\bibinfo {author} {\bibfnamefont {W.}~\bibnamefont
  {Beugeling}}, \bibinfo {author} {\bibfnamefont {E.}~\bibnamefont {Kalesaki}},
  \bibinfo {author} {\bibfnamefont {C.}~\bibnamefont {Delerue}}, \bibinfo
  {author} {\bibfnamefont {Y.-M.}\ \bibnamefont {Niquet}}, \bibinfo {author}
  {\bibfnamefont {D.}~\bibnamefont {Vanmaekelbergh}}, \ and\ \bibinfo {author}
  {\bibfnamefont {C.~M.}\ \bibnamefont {Smith}},\ }\href@noop {} {\bibfield
  {journal} {\bibinfo  {journal} {Nat. Commun.}\ }\textbf {\bibinfo {volume}
  {6}} (\bibinfo {year} {2015})}\BibitemShut {NoStop}%
\bibitem [{\citenamefont {Kalesaki}\ \emph {et~al.}(2014)\citenamefont
  {Kalesaki}, \citenamefont {Delerue}, \citenamefont {Smith}, \citenamefont
  {Beugeling}, \citenamefont {Allan},\ and\ \citenamefont
  {Vanmaekelbergh}}]{kalesaki2014}
  \BibitemOpen
  \bibfield  {author} {\bibinfo {author} {\bibfnamefont {E.}~\bibnamefont
  {Kalesaki}}, \bibinfo {author} {\bibfnamefont {C.}~\bibnamefont {Delerue}},
  \bibinfo {author} {\bibfnamefont {C.~Morais}\ \bibnamefont {Smith}}, \bibinfo
  {author} {\bibfnamefont {W.}~\bibnamefont {Beugeling}}, \bibinfo {author}
  {\bibfnamefont {G.}~\bibnamefont {Allan}}, \ and\ \bibinfo {author}
  {\bibfnamefont {D.}~\bibnamefont {Vanmaekelbergh}},\ }\href@noop {}
  {\bibfield  {journal} {\bibinfo  {journal} {Phys. Rev. X}\ }\textbf {\bibinfo
  {volume} {4}},\ \bibinfo {pages} {011010} (\bibinfo {year}
  {2014})}\BibitemShut {NoStop}
\bibitem [{\citenamefont {K{\'a}lm{\'a}n}\ \emph {et~al.}(2012)\citenamefont
  {K{\'a}lm{\'a}n}, \citenamefont {Kiss}, \citenamefont {Fort{\'a}gh},\ and\
  \citenamefont {Domokos}}]{Kalman2012}
  \BibitemOpen
  \bibfield  {author} {\bibinfo {author} {\bibfnamefont {O.}~\bibnamefont
  {K{\'a}lm{\'a}n}}, \bibinfo {author} {\bibfnamefont {T.}~\bibnamefont
  {Kiss}}, \bibinfo {author} {\bibfnamefont {J.}~\bibnamefont {Fort{\'a}gh}}, \
  and\ \bibinfo {author} {\bibfnamefont {P.}~\bibnamefont {Domokos}},\ }\href
  {http://pubs.acs.org/doi/abs/10.1021/nl203762g} {\bibfield  {journal}
  {\bibinfo  {journal} {Nano Lett.}\ }\textbf {\bibinfo {volume} {12}},\
  \bibinfo {pages} {435} (\bibinfo {year} {2012})}\BibitemShut {NoStop}%
\bibitem [{\citenamefont {G{\"u}nther}\ \emph {et~al.}(2005)\citenamefont
  {G{\"u}nther}, \citenamefont {Kemmler}, \citenamefont {Kraft}, \citenamefont
  {Vale}, \citenamefont {Zimmermann},\ and\ \citenamefont
  {Fort\'agh}}]{Guenther2005a}
  \BibitemOpen
  \bibfield  {author} {\bibinfo {author} {\bibfnamefont {A.}~\bibnamefont
  {G{\"u}nther}}, \bibinfo {author} {\bibfnamefont {M.}~\bibnamefont
  {Kemmler}}, \bibinfo {author} {\bibfnamefont {S.}~\bibnamefont {Kraft}},
  \bibinfo {author} {\bibfnamefont {C.~J.}\ \bibnamefont {Vale}}, \bibinfo
  {author} {\bibfnamefont {C.}~\bibnamefont {Zimmermann}}, \ and\ \bibinfo
  {author} {\bibfnamefont {J.}~\bibnamefont {Fort\'agh}},\ }\href@noop {}
  {\bibfield  {journal} {\bibinfo  {journal} {Phys. {R}ev. {A}}\ }\textbf
  {\bibinfo {volume} {71}},\ \bibinfo {pages} {063619} (\bibinfo {year}
  {2005})}\BibitemShut {NoStop}
\bibitem [{\citenamefont {Stibor}\ \emph {et~al.}(2010)\citenamefont {Stibor},
  \citenamefont {Bender}, \citenamefont {K{\"u}hnhold}, \citenamefont
  {Fort\'agh}, \citenamefont {Zimmermann},\ and\ \citenamefont
  {G{\"u}nther}}]{Stibor2010}
  \BibitemOpen
  \bibfield  {author} {\bibinfo {author} {\bibfnamefont {A.}~\bibnamefont
  {Stibor}}, \bibinfo {author} {\bibfnamefont {H.}~\bibnamefont {Bender}},
  \bibinfo {author} {\bibfnamefont {S.}~\bibnamefont {K{\"u}hnhold}}, \bibinfo
  {author} {\bibfnamefont {J.}~\bibnamefont {Fort\'agh}}, \bibinfo {author}
  {\bibfnamefont {C.}~\bibnamefont {Zimmermann}}, \ and\ \bibinfo {author}
  {\bibfnamefont {A.}~\bibnamefont {G{\"u}nther}},\ }\href
  {http://iopscience.iop.org/1367-2630/12/6/065034} {\bibfield  {journal}
  {\bibinfo  {journal} {New J. Phys.}\ }\textbf {\bibinfo {volume} {12}},\
  \bibinfo {pages} {065034} (\bibinfo {year} {2010})}\BibitemShut {NoStop}%
\bibitem [{\citenamefont {Federsel}\ \emph {et~al.}(2015)\citenamefont
  {Federsel}, \citenamefont {Rogulj}, \citenamefont {Menold}, \citenamefont
  {Fort\'agh},\ and\ \citenamefont {G\"unther}}]{Federsel2015a}
  \BibitemOpen
  \bibfield  {author} {\bibinfo {author} {\bibfnamefont {P.}~\bibnamefont
  {Federsel}}, \bibinfo {author} {\bibfnamefont {C.}~\bibnamefont {Rogulj}},
  \bibinfo {author} {\bibfnamefont {T.}~\bibnamefont {Menold}}, \bibinfo
  {author} {\bibfnamefont {J.}~\bibnamefont {Fort\'agh}}, \ and\ \bibinfo
  {author} {\bibfnamefont {A.}~\bibnamefont {G\"unther}},\ }\href {\doibase
  10.1103/PhysRevA.92.033601} {\bibfield  {journal} {\bibinfo  {journal} {Phys.
  Rev. A}\ }\textbf {\bibinfo {volume} {92}},\ \bibinfo {pages} {033601}
  (\bibinfo {year} {2015})}\BibitemShut {NoStop}
\bibitem [{\citenamefont {Kalman}\ \emph {et~al.}(2016)\citenamefont
  {Kalman}, \citenamefont {Darazs}, \citenamefont {Brennecke},\ and\ \citenamefont {Domokos}}]{Kalman2016}
  \BibitemOpen
  \bibfield  {author} {\bibinfo {author} {\bibfnamefont {O.}~\bibnamefont
  {K\'alm\'an}}, \bibinfo {author} {\bibfnamefont {Z.}~\bibnamefont {Dar\'azs}},
  \bibinfo {author} {\bibfnamefont {F.}~\bibnamefont {Brennecke}}, \ and\ \bibinfo
  {author} {\bibfnamefont {P.}~\bibnamefont {Domokos}},\ }\href {\doibase
  10.1103/PhysRevA.94.033626} {\bibfield  {journal} {\bibinfo  {journal} {Phys.
  Rev. A}\ }\textbf {\bibinfo {volume} {94}},\ \bibinfo {pages} {033626}
  (\bibinfo {year} {2016})}\BibitemShut {NoStop}
\bibitem [{\citenamefont {Vainio}\ \emph {et~al.}(2006)\citenamefont {Vainio},
  \citenamefont {Vale}, \citenamefont {Davis}, \citenamefont {Heckenberg},\
  and\ \citenamefont {Rubinsztein-Dunlop}}]{Vainio2006}
  \BibitemOpen
  \bibfield  {author} {\bibinfo {author} {\bibfnamefont {O.}~\bibnamefont
  {Vainio}}, \bibinfo {author} {\bibfnamefont {C.~J.}\ \bibnamefont {Vale}},
  \bibinfo {author} {\bibfnamefont {M.~J.}\ \bibnamefont {Davis}}, \bibinfo
  {author} {\bibfnamefont {N.~R.}\ \bibnamefont {Heckenberg}}, \ and\ \bibinfo
  {author} {\bibfnamefont {H.}~\bibnamefont {Rubinsztein-Dunlop}},\ }\href
  {\doibase 10.1103/PhysRevA.73.063613} {\bibfield  {journal} {\bibinfo
  {journal} {Phys. Rev. A}\ }\textbf {\bibinfo {volume} {73}},\ \bibinfo
  {pages} {063613} (\bibinfo {year} {2006})}\BibitemShut {NoStop}%
\bibitem [{\citenamefont {Esslinger}\ \emph {et~al.}(2000)\citenamefont
  {Esslinger}, \citenamefont {Bloch},\ and\ \citenamefont
  {H\"ansch}}]{Esslinger2000}
  \BibitemOpen
  \bibfield  {author} {\bibinfo {author} {\bibfnamefont {T.}~\bibnamefont
  {Esslinger}}, \bibinfo {author} {\bibfnamefont {I.}~\bibnamefont {Bloch}}, \
  and\ \bibinfo {author} {\bibfnamefont {T.~W.}\ \bibnamefont {H\"ansch}},\
  }\href {http://www.tandfonline.com/doi/abs/10.1080/09500340008232192}
  {\bibfield  {journal} {\bibinfo  {journal} {J. Mod. Opt.}\ }\textbf {\bibinfo
  {volume} {47}},\ \bibinfo {pages} {2725} (\bibinfo {year}
  {2000})}\BibitemShut {NoStop}
\bibitem [{\citenamefont {Bloch}\ \emph {et~al.}(2000)\citenamefont {Bloch},
  \citenamefont {H{\"a}nsch},\ and\ \citenamefont {Esslinger}}]{Bloch2000}%
  \BibitemOpen
  \bibfield  {author} {\bibinfo {author} {\bibfnamefont {I.}~\bibnamefont
  {Bloch}}, \bibinfo {author} {\bibfnamefont {T.~W.}\ \bibnamefont
  {H{\"a}nsch}}, \ and\ \bibinfo {author} {\bibfnamefont {T.}~\bibnamefont
  {Esslinger}},\ }\href {http://dx.doi.org/10.1038/35003132} {\bibfield
  {journal} {\bibinfo  {journal} {Nature (London)}\ }\textbf {\bibinfo {volume}
  {403}},\ \bibinfo {pages} {166} (\bibinfo {year} {2000})}\BibitemShut
  {NoStop}
\bibitem [{\citenamefont {Bourdel}\ \emph {et~al.}(2006)\citenamefont
  {Bourdel}, \citenamefont {Donner}, \citenamefont {Ritter}, \citenamefont
  {{\"O}ttl}, \citenamefont {K{\"o}hl},\ and\ \citenamefont
  {Esslinger}}]{Bourdel2006}
  \BibitemOpen
  \bibfield  {author} {\bibinfo {author} {\bibfnamefont {T.}~\bibnamefont
  {Bourdel}}, \bibinfo {author} {\bibfnamefont {T.}~\bibnamefont {Donner}},
  \bibinfo {author} {\bibfnamefont {S.}~\bibnamefont {Ritter}}, \bibinfo
  {author} {\bibfnamefont {A.}~\bibnamefont {{\"O}ttl}}, \bibinfo {author}
  {\bibfnamefont {M.}~\bibnamefont {K{\"o}hl}}, \ and\ \bibinfo {author}
  {\bibfnamefont {T.}~\bibnamefont {Esslinger}},\ }\href {\doibase
  10.1103/PhysRevA.73.043602} {\bibfield  {journal} {\bibinfo  {journal} {Phys.
  Rev. A}\ }\textbf {\bibinfo {volume} {73}},\ \bibinfo {pages} {043602}
  (\bibinfo {year} {2006})}\BibitemShut {NoStop}
\bibitem [{\citenamefont {Yarlagadda}(2010)}]{Yarlagadda2010}%
  \BibitemOpen
  \bibfield  {author} {\bibinfo {author} {\bibfnamefont {R.~K.~Rao}\
  \bibnamefont {Yarlagadda}},\ }\href@noop {} {\emph {\bibinfo {title} {Analog
  and digital signals and systems}}},\ Vol.~\bibinfo {volume} {1}\ (\bibinfo
  {publisher} {Springer},\ \bibinfo {year} {2010})\BibitemShut {NoStop}%
\bibitem [{Not()}]{Note1}%
  \BibitemOpen
  \href@noop {} {}\bibinfo {note} {This is because the correlation function has been calculated on the arrival times of the individual counts, and not from the temporally binned detector signal. The correlation function is thus much better suited for extracting the visibility.}\BibitemShut {Stop}
\bibitem [{\citenamefont {{\"O}ttl}\ \emph {et~al.}(2005)\citenamefont
  {{\"O}ttl}, \citenamefont {Ritter}, \citenamefont {K{\"o}hl},\ and\
  \citenamefont {Esslinger}}]{Oettl2005}
  \BibitemOpen
  \bibfield  {author} {\bibinfo {author} {\bibfnamefont {A.}~\bibnamefont
  {{\"O}ttl}}, \bibinfo {author} {\bibfnamefont {S.}~\bibnamefont {Ritter}},
  \bibinfo {author} {\bibfnamefont {M.}~\bibnamefont {K{\"o}hl}}, \ and\
  \bibinfo {author} {\bibfnamefont {T.}~\bibnamefont {Esslinger}},\ }\href
  {\doibase 10.1103/PhysRevLett.95.090404} {\bibfield  {journal} {\bibinfo
  {journal} {Phys. Rev. Lett.}\ }\textbf {\bibinfo {volume} {95}},\ \bibinfo
  {pages} {090404} (\bibinfo {year} {2005})}\BibitemShut {NoStop}
\bibitem [{\citenamefont {Jing}\ \emph {et~al.}(2000)\citenamefont
  {Jing}, \citenamefont {Chen},\ and\
  \citenamefont {Ge}}]{Jing2000}
  \BibitemOpen
  \bibfield  {author} {\bibinfo {author} {\bibfnamefont {Hui}~\bibnamefont
  {Jing}}, \bibinfo {author} {\bibfnamefont {Jing-Lin}~\bibnamefont {Chen}},
  \ and\
  \bibinfo {author} {\bibfnamefont {Mo-Lin}~\bibnamefont {Ge}},\ }\href
  {\doibase 10.1103/PhysRevA.63.015601} {\bibfield  {journal} {\bibinfo
  {journal} {Phys. Rev. A}\ }\textbf {\bibinfo {volume} {63}},\ \bibinfo
  {pages} {015601} (\bibinfo {year} {2000})}\BibitemShut {NoStop}
\bibitem [{\citenamefont {Jing}\ \emph {et~al.}(2001)\citenamefont
  {Jing}, \citenamefont {Chen},\ and\
  \citenamefont {Ge}}]{Jing2001}
  \BibitemOpen
  \bibfield  {author} {\bibinfo {author} {\bibfnamefont {Hui}~\bibnamefont
  {Jing}}, \bibinfo {author} {\bibfnamefont {Jing-Lin}~\bibnamefont {Chen}},
  \ and\
  \bibinfo {author} {\bibfnamefont {Mo-Lin}~\bibnamefont {Ge}},\ }\href
  {\doibase 10.1103/PhysRevA.65.015601} {\bibfield  {journal} {\bibinfo
  {journal} {Phys. Rev. A}\ }\textbf {\bibinfo {volume} {65}},\ \bibinfo
  {pages} {015601} (\bibinfo {year} {2001})}\BibitemShut {NoStop}
\bibitem [{\citenamefont {Haine}\ \emph {et~al.}(2005)\citenamefont
  {Haine},\ and\
  \citenamefont {Hope}}]{Haine2005}
  \BibitemOpen
  \bibfield  {author} {\bibinfo {author} {\bibfnamefont {S. A.}~\bibnamefont
  {Haine}},
  \ and\
  \bibinfo {author} {\bibfnamefont {J. J.}~\bibnamefont {Hope}},\ }\href
  {\doibase 10.1103/PhysRevA.72.033601} {\bibfield  {journal} {\bibinfo
  {journal} {Phys. Rev. A}\ }\textbf {\bibinfo {volume} {72}},\ \bibinfo
  {pages} {033601} (\bibinfo {year} {2005})}\BibitemShut {NoStop}
\bibitem [{\citenamefont {Nazarov}(2009)}]{Nazarov2009}
  \BibitemOpen
  \bibfield  {author} {\bibinfo {author} {\bibfnamefont {Yuli~V.}\
  \bibnamefont {Nazarov}}\ and\ \bibinfo {author} {\bibfnamefont {M.}~\bibnamefont {Yaroslav}},\ }\href@noop {} {\emph {\bibinfo {title} {Quantum transport: introduction to nanoscience}}}\ (\bibinfo
  {publisher} {Cambridge University Press},\ \bibinfo {year} {2009})\BibitemShut {NoStop}
\end{thebibliography}
\end{document}